\documentstyle[12pt,titlepage]{article}
\newcommand{\seqnoll}{\setcounter{equation}{0}}

\newcommand{\Eqref}[1]{Eq.(\ref{#1})}

\newcommand{\bea}{\begin{eqnarray}}
\newcommand{\eea}{\end{eqnarray}}
\newcommand{\be}{\begin{equation}}
\newcommand{\ee}{\end{equation}}
\newcommand{\bc}{\begin{center}}
\newcommand{\ec}{\end{center}}
\newcommand{\ba}{\begin{array}}
\newcommand{\ea}{\end{array}}
\newcommand{\btab}{\begin{tabular}}
\newcommand{\etab}{\end{tabular}}
\newcommand{\bfig}{\begin{figure}}
\newcommand{\efig}{\end{figure}}
\newcommand{\non}{\nonumber}

\newcommand{\BOX}{\hbox {$\sqcap$ \kern -1em $\sqcup$}}

\newcommand{\real}{{\rm I \! R}}

\newcommand{\complex}{{\rm C\!\!\! I\,\,}}
\def\bigid{\leavevmode\hbox{\small1\kern-3.8pt\normalsize1}}
\def\id{\leavevmode\hbox{\small1\kern-3.3pt\normalsize1}}

\newcommand{\Abs}[1]{\left|#1\right|}

\newcommand{\inv}[1]{\frac{1}{#1}}

\newcommand{\ra}{\rangle}
\newcommand{\la}{\langle}

\newcommand{\Tr}{{\rm Tr}}
\newcommand{\pol}{{\rm pol}}

\newcommand{\bt}{\tilde{\beta}}

\newcommand{\advp}[3]{{\it  Adv. in Phys. }{{\bf #1} {(#2)} {#3}}}

\newcommand{\annp}[3]{{\it  Ann. Phys. (N.Y.) }{{\bf #1} {(#2)} {#3}}}
\newcommand{\cmp}[3]{{\it  Comm. Math. Phys.} {{ \bf #1} {(#2)} {#3}}}

\newcommand{\jmp}[3]{{\it  J. Math. Phys.} {{ \bf #1} {(#2)} {#3}}}

\newcommand{\lmp}[3]{{\it Lett. Math. Phys. }{{\bf #1} {(#2)} {#3}}}
\newcommand{\mpl}[3]{{\it  Mod. Phys. Lett. }{{\bf #1} {(#2)} {#3}}}

\newcommand{\np}[3]{{\it  Nucl. Phys. }{{\bf #1} {(#2)} {#3}}}
\newcommand{\pr}[3]{{\it Phys. Rev.}{{ \bf #1} {(#2)} {#3}}}

\newcommand{\prl}[3]{ {\it Phys. Rev. Lett.}{{ \bf #1} {(#2)} {#3}}}
\newcommand{\pl}[3]{{\it  Phys. Lett. }{{\bf #1} {(#2)} {#3}}}
\newcommand{\prep}[3]{{\it Phys. Rep. }{{\bf #1} {(#2)} {#3}}}

\newcommand{\ptp}[3]{{\it  Prog. Theor. Phys. }{{\bf #1} {(#2)} {#3}}}

\newcommand{\phys}[3]{{\it Physica }{{\bf #1} {(#2)} {#3}}}

\newcommand{\caa}{{${\cal A}_{a}\ $}}
\newcommand{\cax}{{${\cal A}_\xi\ $}}
\newcommand{\cfa}{{${\cal F}_{a}\ $}}
\newcommand{\cfx}{{${\cal F}_\xi\ $}}

\newcommand{\at}{\tilde{a}}
\newcommand{\ad}{{a^\dagger}}
\newcommand{\atd}{{\tilde{a}^\dagger}}
\renewcommand{\bt}{\tilde{b}}
\newcommand{\bd}{{b^\dagger}}
\newcommand{\btd}{{\tilde{b}^\dagger}}
\newcommand{\xt}{{\tilde{\xi}}}
\newcommand{\xd}{{\xi^\dagger}}
\newcommand{\xtd}{{\tilde{\xi}^\dagger}}
\newcommand{\et}{{\tilde{\eta}}}
\newcommand{\ed}{{\eta^\dagger}}
\newcommand{\etd}{{\tilde{\eta}^\dagger}}
\newcommand{\avacl}{{\la {\cal O}_a|}}
\newcommand{\xvacl}{{\la {\cal O}_\xi|}}
\newcommand{\avacr}{{| {\cal O}_a\ra}}
\newcommand{\xvacr}{{| {\cal O}_\xi\ra}}
\newcommand{\bvacr}{{| {\cal O}(\beta)\ra}}
\normalsize

\begin{document}
%
\large
\thispagestyle{empty}
\begin{flushright} NORDITA--93/33   P \\
                   hep-ph/9304089  \\
                   March 1993  \end{flushright}
\bc
\normalsize
{\LARGE\bf Generalizations of the thermal \vspace{4mm}\\
      Bogoliubov transformation}
\ec
\vspace*{1cm}
\bc
{\large P. Elmfors} \\
\normalsize
NORDITA, Blegdamsvej 17\\
 DK--2100 K\o benhavn \O, Denmark\\
E--mail: elmfors@nordita.dk
\ec
\vspace*{8mm}
\bc
{\large H. Umezawa} \\
\normalsize
The Theoretical Physics Institute\\
University of Alberta, Edmonton\\
Alberta T6G 2J1, Canada\\
E--mail: umwa@phys.ualberta.ca
\ec
\vspace*{2cm}
\bc
{\bf Abstract} \\
\ec
{\normalsize
The thermal Bogoliubov transformation in thermo field dynamics
is generalized in two respects. First, a generalization of the
$\alpha$--degree of freedom to tilde non--conserving representations
is considered. Secondly, the usual $2\times2$ Bogoliubov matrix
 is extended to a $4\times4$ matrix including mixing of modes with
non--trivial multiparticle correlations. The analysis is carried
out for both bosons and fermions.}
\newpage
%
\normalsize
\setcounter{page}{1}
\section{Introduction}
\seqnoll
\label{intro}
In recent years there have been two directions of development
of the real time formulation of quantum field theory at finite
temperature called thermo field dynamics (TFD). One problem
has been to reconcile the calculations of vertex functions with
the result from the imaginary time formalism
\cite{Kobes9091,AurencheB92}. Another important issue has been
to generalize the standard TFD to non--equilibrium systems,
in particular to time dependent situations
\cite{UmezawaY90,UmezawaY92,YamanakaUNA92,NakamuraUY92}.
In connection with the second problem one has been led to
consider generalizations of the standard thermal Bogoliubov
transformation. More precisely, there is a freedom of
choosing the parameters in the Bogoliubov matrix corresponding
to a given density matrix. That freedom was used in
Ref.\cite{EvansHUY92} to simplify the calculation of expectation
values in the interaction picture. Only two choices give
time and anti--time ordered products for the perturbation
expansion. Furthermore, the Boltzmann equation in
\cite{YamanakaUNA92} gives increasing entropy only with the
choice $\alpha=1$ ($\alpha$ is one of the parameters in the
Bogoliubov transformation), and when the thermal state is given at
$t_0=-\infty$. The choice $\alpha=0$ and $t_0=\infty$
leads to decreasing entropy.
In the general case the density matrix, or
the thermal state, may be given at
some finite time. The corresponding Bogoliubov transformation
can also be highly non--linear to include the initial
correlation of particles.

In this paper we approach the problem by studying all
possible choices of a linear Bogoliubov transformation
for a given equilibrium density matrix, and discuss
the properties under Hermitian and tilde conjugation
for different choices (see Sec.\ref{genalpha}).
Non--linear Bogoliubov transformations are in general
very difficult to deal with but we can quite easily extend
the usual $2\times2$ thermal matrix to a $4\times4$
matrix
in several ways. In Sec.\ref{fbyfbose} we study
some relevant extensions for bosons. The whole discussion
is repeated for  fermions in Sec.\ref{fbyffermi}.

%
\section{Generalizations of the $\alpha$--degree of freedom}
\seqnoll
\label{genalpha}
There has recently been much discussion about the
freedom of choosing the  thermal
vacuum in TFD corresponding to a given density matrix
\cite{UmezawaY90,EvansHUY92,HenningU92}. In particular
the cyclicity of
the trace gives the $\alpha$--degree of freedom where $\alpha$
is defined by $\la A\ra =
\Tr (\rho^{1-\alpha}A\rho^{\alpha}) =
\la \rho^{1-\alpha}|A|\rho^{\alpha}\ra$. The last
expectation value is calculated in the
Hilbert space of density matrices and the scalar
product is defined by the trace \cite{HaagHW67}.
Note that $\la \rho^{\alpha}|$
is {\it not} the dual
of
$|\rho^{1-\alpha}\ra$.
In TFD the two vectors are constructed from the zero temperature
vacuum $|0,\tilde{0}\ra$ as
\be
|\rho^{\alpha}\ra=\sum_{m,n} \la n|\rho^{\alpha}|m\ra
|m,\tilde{n}\ra \ ,\quad\quad  \\
\la\rho^{1-\alpha}|=\sum_{m,n} \la m|\rho^{1-\alpha}|n\ra
\la m,\tilde{n}| \ .
\ee

In order to keep clear the meaning of Hermiticity, unitarity,
bra-- and ket--vectors, and tilde
conjugation etc., we define two distinct Hilbert
(Fock) spaces \cfa and \cfx. They are
generated from vacuum vectors $\avacr$
and $\xvacr$ by acting with the creation operators ($\ad,\atd$) and
($\xd,\xtd$) respectively. The two spaces are trivially isomorphic.
In each space Hermitian and tilde conjugation are defined
in the usual way and when an
operator is said to be e.g. unitary it is meant to
be with respect to the structure in the space
it is acting. The operators ($a,\atd,\ad,\at$) and
($\xi,\xtd,\xd,\xt$) satisfy the usual canonical
commutation relations (CCR)
and we call the algebras that they generate \caa and \cax.

We are now interested in mappings $\phi$:\ \caa
$\mapsto$ \cax other than the trivial one. In
particular, mappings that mix tilde
and non--tilde operators are interesting since they
represent mixed states. Then thermal expectation
values of polynomials in $a$ and $\ad$ ($\pol(a,\ad)$)
are computed by
\be
\Tr(\rho\, \pol(a,\ad))= \xvacl \phi(\pol(a,\ad))\xvacr\ .
\ee
The two vacuum vectors $\avacr$ and
$\xvacr$ shall be thought of as the zero and finite
temperature vacuum vectors. In this paper we only consider
linear mappings though non--linear mappings are required if
general density matrices should be represented in this way.
Even in such cases the linear mappings are frequently useful
when the non--linear effects can be treated by perturbation
calculation.
\\ \\
We shall see how the $\alpha$--degree of freedom
emerges together
with other parameters when we consider
non--trivial mappings between \cfa and \cfx.
The most commonly used Bogoliubov transformation in TFD is
\be
\label{standardbtrf}
\left(\ba{c} a \\ \atd \ea\right)\stackrel{\phi}{\longmapsto}
\left(\ba{cc}\cosh\theta & \sinh\theta \\
 \sinh\theta & \cosh\theta \ea\right)
\left(\ba{c} \xi \\ \xtd \ea\right)
\equiv B_2(\theta)
\left(\ba{c} \xi \\ \xtd \ea\right)\ ,
\ee
and the Hermitian and tilde conjugates are defined anti--linearly.
This represents the finite temperature density matrix
\be
\rho=\frac{\exp(-\beta\omega\,\ad a)}
{\Tr(\exp(-\beta\omega\, \ad a))}\ ,
\ee
where the inverse temperature $\beta$ is related to $\theta$
through
\be
\sinh^2\theta=\inv{e^{\beta\omega}-1}\ .
\ee
\\ \\
In this section we restrict ourselves to mappings that mix
 $a$ and $\atd$, and their Hermitian
conjugates. The most general linear mapping is
\be
\label{mostgen}
\phi:\ a^\mu \stackrel{\phi}{\longmapsto}
B^{\mu\nu}\xi^\nu \ ,\
\ad^\mu \stackrel{\phi}{\longmapsto}
C^{\mu\nu} \xi^\nu\ .
\ee
In the following we use the doublet notation, i.e.
$a^\mu=(a,\atd)$ and $\ad^\mu=(\ad,\at)$. We require the
CCR to be
conserved under the mapping $\phi$ (we denote the Pauli
matrices by $\tau_i$),
\be
\label{CCRcond}
[a^\mu, \ad^\nu]=\tau_3^{\mu\nu}=(B\tau_3 C^T)^{\mu\nu}
=[\xi^\mu,\xd^\nu]\ ,
\ee
so the only condition on $B\in GL(2,\complex)$ is
that it should be invertible to define $C$. This
mapping leaves the free Hamiltonian invariant,
\be
\hat{H}(a)=\at a-\atd\at\ \stackrel{\phi}{\longmapsto}\
\hat{H}(\xi)=\xd\xi-\xtd\xt\ .
\ee
There are eight independent real parameters in $B$
(except for the condition $\det(B)\neq 0$)
but only one combination appears  in the expectation
values of observables. For example, the particle
number  $\ad a$ is
\be
\xvacl\phi(\ad a)\xvacr =
\frac{\beta\gamma}{\zeta\delta-\beta\gamma}\ ,
\ee
if we parametrize $B$ by
\be
\label{B}
B=\left(\ba{cc} \zeta & \beta \\ \gamma & \delta \ea\right)\ .
\ee
It turns out that the only physical parameter is
$\beta\gamma/\zeta\delta$, which should be chosen in
$[0,1]\in\real$ since the expectation value of the number
operator must be real and positive.
\\ \\
The most general mapping conserves CCR but not
Hermiticity and tilde conjugation in the
sense that
\be
\phi(\ad)\neq\phi(a)^\dagger \ ,
\ \phi(\at)\neq\widetilde{\phi(a)}\ .
\ee
Note that $\dagger$ and $\sim$ refer
to different Hilbert spaces depending on which
operator they act. In particular that means that
the image of an Hermitian operator is not
necessarily Hermitian. When the Hamiltonian is expressed
in terms of $\xi$--operators it may be
non--Hermitian and the time evolution non--unitary in \cfx.
\\ \\
In many recent publications the mapping
has been required to preserve the tilde  but not Hermitian
conjugation
\cite{ArimitsuUY87,UmezawaY88,EvansHUY90,UmezawaY90,HenningU92}.
Since $a^\mu=\tau_1^{\mu\nu}\tilde{a}^{\dagger\nu}$, the first relation in
\Eqref{mostgen} leads to
$\tilde{a}^{\dagger\mu}\mapsto(\tau_1B\tau_1)^{\mu\nu}
\tilde{\xi}^{\dagger\nu}$. On the other hand the tilde
conjugation of the
second relation in \Eqref{mostgen} gives
$\tilde{a}^{\dagger\mu}\mapsto(C^*)^{\mu\nu}
\tilde{\xi}^{\dagger\nu}$. The invariance under tilde conjugation means
$C^*=\tau_1B\tau_1$. This, together with \Eqref{CCRcond},
gives
\be
B\tau_2 B^\dagger=\tau_2\ .
\ee
The solution is
\be
B=e^{i\psi}\left(\ba{cc} \zeta & \beta \\
\gamma & \delta \ea\right)\ ,
\ee
where $\zeta ,\  \beta , \  \gamma,\ \delta,\ \psi\in\real$,
 $\psi$ is an arbitrary phase and
$\zeta\delta-\beta\gamma=1$. Thus $B\in SL(2,\real)\times U(1)$.

The commonly used tilde--preserving Bogoliubov transformation is
\cite{HenningU92,Umezawa93}
\be
B=\inv{\sqrt{1-f}}\left(\ba{cc} e^{-s} & f^\alpha e^s \\
            f^{1-\alpha}e^{-s} & e^s \ea\right)\ .
\ee
(Note that in Refs.\cite{HenningU92,Umezawa93} this matrix is
called $B^{-1}$.)
Then the relation between ($\zeta,\beta,\gamma,\delta$) and
the parameters ($\alpha,s,f$) is
\be
f=\frac{\beta\gamma}{\zeta\delta}\ , \
s=\inv{2}\log(\frac{\delta}{\zeta})
\ , \ \alpha=\frac{\log(\beta\zeta)}
{\log(\beta\gamma)-\log(\zeta\delta)}\ .
\ee
The only combination that occurs in physical expectation values is
\be
n=\frac{f}{1-f}=\beta\gamma\ .
\ee
\\ \\
It is also possible to preserve the Hermitian
conjugation but not tilde conjugation and
that imposes the condition
\be
B\tau_3B^\dagger=\tau_3\ .
\ee
The solution is
\be
\label{Ttrf}
B=e^{i\psi}\left(\ba{cc} \zeta e^{i\phi} &
\beta e^{i\varphi}
\\ \beta e^{-i\varphi} &  \zeta e^{-i\phi} \ea\right)\ ,
\ee
where $\zeta,\beta,\phi,\varphi,\psi\in\real$
and $\zeta^2-\beta^2=1$. This mapping leaves
the Hamiltonian Hermitian so the time evolution in \cfx is unitary.
\\ \\
Finally we can consider mappings that preserve
both Hermiticity and tilde conjugation.
The Bogoliubov transformation is then reduced to the usual
\be
\label{BHT}
B\equiv e^{i\psi}B_2(\theta)=e^{i\psi}
\left(\ba{cc}\cosh\theta & \sinh\theta \\
 \sinh\theta & \cosh\theta \ea\right)\ .
\ee
The phase $e^{i\psi}$ is just the arbitrary
of the phase of a state in \cfx.

Among all the parameters only one combination has physical
meaning for each mapping $\phi$, the others being similar to
the $\alpha$--degree of freedom. For a given value
of e.g. $\xvacl \phi(\ad a)\xvacr$, which fixes
the physical parameter, all mappings considered
so far represent the same density matrix.

The fact that we only consider linear mappings from
$(a,\atd)$ to $(\xi,\xtd)$ implies that multiparticle
expectation values are given in a simple way from
the single particle number. For example, with the
transformation in \Eqref{BHT} we have
\be
\label{novev}
\ba{rcl}
\xvacl \phi(\ad a) \xvacr &=& \sinh^2\theta \ ,\\
\xvacl \phi((\ad)^2 (a)^2) \xvacr &=& 2\sinh^4\theta\ .
\ea
\ee
These expectation values probe different components
of the thermal vacuum belonging to different $n$--particle
subspaces of the Fock space and their values need not be
related for a general density matrix.

In all cases above the free Hamiltonian is left
invariant and satisfies the usual conditions
under Hermitian and tilde conjugation in \cfx.
However, when interactions are included
new terms appear that destroy these properties.
\\ \\
The free propagator for a real field
\be
\phi(x)=\int\frac{d^3k}{\sqrt{(2\pi)^3 2\omega_k}}
\left(\ad_k e^{i(\omega_k t-kx)}+a_k e^{-i(\omega_k t-kx)}\right)\ ,
\ee
in the state $\xvacr$ can be calculated in a
straightforward manner. We find
\bea
D^{ab}(k_0,\vec{k})&=& \left(\ba{cc} \Delta_F & 0 \\
      0 & -\Delta^*_F \ea\right) \non\\
&-& \frac{2\pi i\delta(k_0^2-\omega_k^2)}{\zeta\delta-\beta\gamma}
\left(\ba{cc} \beta\gamma & \beta\delta\theta(k_0)+
\zeta\gamma\theta(-k_0) \\ \zeta\gamma\theta(k_0)+
\beta\delta\theta(-k_0) & \beta\gamma\ea\right)
\eea
where
\be
\Delta_F=\inv{k_0^2-\omega_k^2+i\epsilon}\ .
\ee
It reduces to the well--known expression when the parameters
($\zeta,\beta,\gamma,\delta$) are constrained by preservation of
$\dagger$ and $\sim$ (\Eqref{BHT}).
%
%
\subsection{Another way of describing the
non--standard Bogoliubov transformations}
It is often convenient to formally think of
$\avacr$ and $\xvacr$ as vectors in the
same Hilbert space \cfa. This is not mathematically
rigorous in field theory since the Hilbert spaces
\cfa and \cfx are unitarily inequivalent.
 Any density matrix $\rho$
can be represented by a vector $\xvacr\in$\cfa that
is obtained by a unitary transformation of $\avacr$.
It is usually written
\be
\xvacr=e^{iG}\avacr=U\avacr\ ,
\ee
where $G$ is Hermitian and $U$ is unitary.
It is also possible to choose $U$ to
be tilde invariant $\tilde{U}=U$ which fixes $U$ completely.
The $\xi$ operators that annihilates $\xvacr$ is
obtained by the corresponding transformation
\be
\xi=UaU^{-1}=UaU^\dagger\ .
\ee
Since since $U$ is both unitary and tilde invariant we also have
\be
\xt=U\at U^\dagger\ , \ {\rm and} \ \xd=U\ad U^\dagger\ .
\ee
In the particular case of a linear transformation
between $a$ and $\xi$, $G$ takes the form
\be
G=\psi(\ad a -\atd\at)+i\theta(a\at-\ad\atd)\ ,
\ee
where $\psi$ and $\theta$ are the parameters occuring in \Eqref{BHT}.
\\ \\
The generalizations of the thermal Bogoliubov
transformation considered so far amounts
to replacing $U$ by an
another invertible operator $V$ which may be non--unitary
and non--tilde invariant. Let us first
say a few words about the tilde conjugation. It can be
described by the Tomita--Takesaki modular operator $J_a$
\cite{BratteliR87}
which is defined relative to the vacuum $\avacr$
and the algebras generated by $\{ a,\ad\}$ and $\{ \at,\atd\}$.
It leaves $\avacr$ invariant and maps
$\{ a,\ad\}$ to $\{ \at,\atd\}$ by $J_a aJ_a=\at$ etc.
Given another vacuum $\xvacr$ there is another
$J_\xi$ defined by $J_\xi=VJ_aV^{-1}$. If $V$
is tilde invariant we have $J_\xi=J_a$, otherwise
the tilde conjugation is not preserved. It
is however replaced by a new tilde conjugation rule
relative to the $\xi$--operators and $\xvacr$.

In more physical terms we can say that there is a tilde
conjugation defined for the zero temperature vacuum.
If the transformation to the thermal vacuum commutes with
that tilde conjugation we will have the same tilde
conjugation operator for the thermal vacuum. Otherwise
we have to define a new tilde conjugation with respect
to the thermal vacuum. Note that we can always
find such a new tilde conjugation rule.
\\ \\
Let us consider some general transformation of $\avacr$
\be
\xvacr=V\avacr\ .
\ee
In order that $\xi$ annihilates $\xvacr$ we take
$\xi=VaV^{-1}$. To preserve the CCR we define
$\xi^\ddagger=V\ad V^{-1}$ to be the canonical
conjugate of $\xi$. Note that $\xi^\ddagger\neq\xd$
if $V$ is non--unitary.
The bra state annihilated by $\xi^\ddagger$ is
$\la{\cal O}_\xi^\ddagger|=\avacl V^{-1}$ which
is {\it not} the dual of $\xvacr$ if $V$ is non--unitary.
We also want to define a tilde conjugate of $\xi$,
call it $\check{\xi}$, such that $\xi$ and $\check{\xi}$
commute. The obvious choice is
$\check{\xi}=V\at V^{-1}=J_\xi \xi J_\xi$,
but we note that $\check{\xi}\neq\xt$ if $V\neq\tilde{V}$.
\\ \\
The idea with the usual thermal Bogoliubov
transformation is that thermal expectation
values of observables like polynomials in $a$ and $\ad$
($\pol(a,\ad)$) are obtained
as vacuum expectation values in the thermal vacuum
$\xvacr$. With the transformation V the thermal
expectation value is instead computed as
\be
\Tr(\rho\ \pol(a,\ad))=\la{\cal O}_\xi^\ddagger|
\pol(a,\ad)\xvacr\ .
\ee
We also notice that $\xvacr$ is no longer normalized
if $V\neq V^\dagger$ but satisfies
$\la{\cal O}_\xi^\ddagger\xvacr=1$. In particular, in
the $\alpha=1$ representation of a single oscillator
\cite{EvansHUY92,HenningU92}
at equilibrium we have
\be
\la{\cal O}_\xi^\ddagger|=\sum_n\la n,\tilde{n}|\ ,
\ee
 which is not normalizable. The linear mapping in
\Eqref{B} is still well--defined if, for instance,
$\zeta=\gamma=\delta$ for which $s=0$, $\alpha=1$
and $f=(\zeta^2-1)/\zeta$.

The freedom of choosing different rules for
tilde conjugation arises from the independence
of the unitary operator $U$ in the expectation value
$\Tr(U^\dagger\rho^{1/2}A\rho^{1/2}U)=\la A\ra$.
It has been studied in detail as a gauge degree of
freedom when representing $\rho$ in as states in \cfa
\cite{Uhlmann8690,HenningGM92}.
%
%
%
\section{Thermal and particle mixing Bogoliubov transformations}
\seqnoll
\label{fbyfbose}
The equilibrium state of a harmonic oscillator can
be represented by the simple mapping in \Eqref{BHT}.
It describes independent particles with all
multiparticle correlations determined in terms
of the number expectation value $\xvacl \ad a\xvacr$.
\footnote{In the rest of the paper we shall write
$\xvacl \ad a\xvacr$ as a simplified notation for
$\xvacl \phi(\ad a)\xvacr$.}
This is clearly a very particular state.

In field theory the system is often assumed to be
at equilibrium with respect to the interacting
Hamiltonian at $t=-\infty$ and the vacuum $\bvacr$
satisfies $\hat{H}\bvacr=0$. To do perturbation
theory we rather want to use the thermal vacuum of
the free Hamiltonian $\bvacr_0$ satisfying $\hat{H}_0\bvacr_0=0$.
The difference between the two states is related to
the vertical integration paths in the path
integral formulation of TFD. These paths can
often (but not always \cite{MatsumotoUW86})  be neglected
when $t\rightarrow\pm\infty$, due to clustering
properties of QFT. If, on the other hand, the initial
state is given at some finite time one may have to
deal with non--trivial correlations. The thermal
Bogoliubov transformation is then
non--linear in general.
\\ \\
To do simple things first we shall consider the
kind of states that can be described by generalizing
the  Bogoliubov transformation to a mixing of four operators
while still being linear. Common to these transformations
is that the free Hamiltonian is no longer invariant.
But there is no reason why the initial correlation of
particles should be constrained by the Hamiltonian which
governs the dynamics of the particles.

We shall consider three different kinds of particle
mixing. They are all combinations of some zero and
finite temperature transformation.

First, the simple mixing of $a$ and $\atd$ can be
extended to include $\at$ and $\ad$ as well. This is
a toy model for a state out of equilibrium where the
time dependence of correlation function can be calculated
explicitly for a free Hamiltonian.

Secondly, we study a mixing of two modes, $a$ and $b$,
which can be different momentum modes for instance.
Such a mixing occurs naturally at local equilibrium
where not only two but all modes mix
\cite{ElmforsS92,NakamuraUY92}.

As a third example  a mixing of the superfluidity type between
$a$ and $\bd$ is considered.
\\ \\
The two last cases have in common that the most
general transformation respecting $\dagger$-- and
$\sim$--conjugation has more parameters than
a simple tensor product of zero and finite temperature
transformations. The parameters can be interpreted
as different temperatures for the different modes
and a rotation, but there is also a fourth parameter that
has not been discussed earlier.
\\ \\
{\bf Case I}
\\
We introduce a quadruple notation for the operators
\be
\label{quadruple1}
\xi^\mu=(\xi,\xtd,\xd,\xt)\ ,\ a^\mu=(a,\atd,\ad,\at)\ ,
\ee
with which we can write the CCR as
\be
[\xi^\mu,\xi^\nu]=(\tau_2\otimes\tau_3)^{\mu\nu}\ .
\ee
With this notation the Bogoliubov transformation
is $a=B_4\xi$, where $B_4$ is a $4\times 4$ matrix.
The conditions that CCR, $\dagger$-- and $\sim$--conjugations
are preserved are
\be
\ba{rrcl}
{\rm CCR:} & B_4(\tau_2\otimes\tau_3)B_4^T &=&
      (\tau_2\otimes\tau_3)\ ,\\
\dagger: & B_4^*(\tau_1\otimes\id) &=& (\tau_1\otimes\id)B_4\ , \\
\sim: & B_4^*(\tau_1\otimes\tau_1) &=& (\tau_1\otimes\tau_1)B_4\ .
\ea
\ee
The solution is of the form
\be
B_4=\left(\ba{cc} \alpha & \beta \\
\beta^* & \alpha^* \ea\right)\ ,
\ee
where $\alpha$ and $\beta$ are $2\times2$ matrices
\be
\alpha=\left(\ba{cc} \rho_1 & \rho_2 \\ \rho_2 & \rho_1 \ea\right)
e^{i\psi_1}\ ,\
\beta=\left(\ba{cc} r_1 & r_2 \\ r_2 & r_1 \ea\right)e^{i\psi_2}\ ,
\ee
with the conditions that ($r_i,\rho_i,\psi_i\in\real$)
\be
\ba{rcl}
(\rho_1^2+r_2^2)-(\rho_2^2+r_1^2)&=& 1\ , \\
\rho_1r_2-\rho_2r_1&=&0\ .
\ea
\ee
If we assume that $\rho_1^2-\rho_2^2\geq1$ we can
satisfy the conditions by parametrizing $B_4$ as a
simple tensor product
\be
B_4=\left(\ba{rr}  e^{i\psi_1}\cosh\phi &  e^{i\psi_2}\sinh\phi\\
 e^{-i\psi_2}\sinh\phi &  e^{-i\psi_1}\cosh\phi\ea\right)\otimes
B_2(\theta)\ .
\ee
\\ \\
The free Hamiltonian, expressed in terms of $\xi$--operators,
is not invariant under this transformation and it does
not annihilate the vacuum.
\bea
\hat{H}_0(a)&=&\omega(\ad a-\atd\at)\mapsto\hat{H}(\xi)
=\omega(1+2\sinh^2\phi)
(\xd\xi-\xtd\xt) \non\\
& &+\, \omega\frac{\sinh 2\phi}{2}
[e^{i(\psi_1-\psi_2)}(\xi\xi-\xtd\xtd) +
e^{-i(\psi_1-\psi_2)}(\xd\xd-\xt\xt)]\ .
\eea
This means that the thermal vacuum is not stationary.
We note that $\hat{H}(\xi)$ satisfy the usual conditions
\be
\hat{H}^\dagger=\hat{H}\ ,\ \widetilde{\hat{H}}=-\hat{H}\ .
\ee
The expectation value of the number operator is given by
\be
\xvacl \ad a\xvacr = \cosh^2\theta\sinh^2\phi+
      \sinh^2\theta\cosh^2\phi\ ,
\ee
and
\be
\xvacl(\ad)^2(a)^2\xvacr= 2\xvacl \ad a\xvacr^2+
      \cosh^2\phi\sinh^2\phi\
      (\sinh^2\theta+\cosh^2\theta)^2\ ,
\ee
which, by comparing with \Eqref{novev}, clearly shows that there
is a non--trivial multiparticle correlation.
\\ \\
{\bf Case II}
\\
In the case of mode mixing we include two independent
operators $a$ and $b$ in the quadruple
\be
a^\mu=(a,\atd,b,\btd)\ ,\quad \xi^\mu=(\xi,\xtd,\eta,\etd)\ .
\ee
They can, for instance, be modes of different momenta
\cite{ElmforsS92,NakamuraUY92}.
The Bogoliubov transformation $a=B_4\xi$ preserves
CCR, $\dagger$-- and $\sim$--conjugation if it satisfies
\be
\label{bc2}
\ba{rcl}
B_4(\id\otimes\tau_3)B_4^\dagger &=& (\id\otimes\tau_3)\ ,\\
B_4(\id\otimes\tau_1) &=& (\id\otimes\tau_1)B_4\ .
\ea
\ee
A tensor product of  a $SU(2)$ and a $SO(1,1)$ rotation
satisfies \Eqref{bc2} but it is not the most general solution.
The most general complex matrix satisfying \Eqref{bc2} is
rather complicated so we restrict it to be real for
simplicity. We then have
\be
B_4=\left(\ba{cc} \alpha & \beta
\\ \gamma & \delta \ea\right)\ ,\
\alpha=\left(\ba{cc} \alpha_1 & \alpha_2
\\ \alpha_2 & \alpha_1 \ea\right)
{\rm etc. \ for\ }\beta,\gamma\ {\rm and\ }\delta\ ,
\ee
with the conditions
\be
\ba{rcl}
(\alpha_1^2-\alpha_2^2)+(\beta_1^2-\beta_2^2)&=&1\ , \non\\
(\gamma_1^2-\gamma_2^2)+(\delta_1^2-\delta_2^2)&=&1\ , \non\\
\alpha_1\gamma_1-\alpha_2\gamma_2+
\beta_1\delta_1-\beta_2\delta_2&=&0\ , \non\\
\alpha_1\gamma_2-\alpha_2\gamma_1+
\beta_1\delta_2-\beta_1\delta_2&=&0\ .
\ea
\ee
If we assume that $(\alpha_1^2-\alpha_2^2),
(\beta_1^2-\beta_2^2), (\gamma_1^2-\gamma_2^2),
(\delta_1^2-\delta_2^2)\geq 0$ it is easy to
parametrize the solutions as
\be
\label{B4solII}
B_4=\left(\ba{cc} B_2(\theta_1)\cos\phi &
B_2(\theta_2-\delta)\sin\phi \\
\mp B_2(\theta_1+\delta)\sin\phi &
\pm B_2(\theta_2)\cos\phi \ea\right)\ .
\ee
The two parameters $\theta_1$ and $\theta_2$ are
interpreted as the Bogoliubov parameters referring
to the temperatures of the $a$ and $b$ particles, which
may be different. The $\phi$ parameter is the mixing angle
between $a$ and $b$. Finally, there is fourth parameter $\delta$
which only exist when the particle and thermal mixing
occur simultaneously.

The free Hamiltonian for two modes, when the upper sign
in \Eqref{B4solII} is chosen, is transformed into
\bea
\label{HamboseII}
\hat{H}_0(a,b)&=&\omega_a(\ad a-\atd\at)+\omega_b(\bd b-\btd\bt)\non\\
      \mapsto\hat{H}(\xi,\eta)&=&(\omega_a\cos^2\phi+\omega_b\sin^2\phi)
      (\xd\xi-\xtd\xt)+(\omega_b\cos^2\phi+\omega_a\sin^2\phi)
      (\ed\eta-\etd\et)\non\\
      & &+
      \frac{\sin 2\phi}{2}(\omega_a-\omega_b)\left\{
      \cosh(\theta_1-\theta_2+\delta)
      [\xd\eta+\ed\xi-\xtd\et-\etd\xt] \right. \non\\
      & &+ \left.
      \sinh(\theta_1-\theta_2+\delta)
      [\eta\xt+\ed\xtd-\xi\et-\xd\etd]\right\}\ .
\eea
It turns out that the Hamiltonian depends only on $\phi$ and the
combination $\theta_1-\theta_2+\delta$.
Other operators may depend on other combinations and,
for instance, the number expectation values for the two modes are
\bea
\xvacl\ad a\xvacr &=& \cos^2\phi\sinh^2\theta_1+
      \sin^2\phi\sinh^2(\theta_2-\delta)\ ,\non\\
\xvacl\bd b\xvacr &=& \cos^2\phi\sinh^2\theta_2+
      \sin^2\phi\sinh^2(\theta_1+\delta)\ .
\eea
\\ \\
{\bf Case III}
\\
A mixing between $a$ and $\bd$ is interesting in the theory
of superfluid bosons where $a=a_k$ and $\bd=\ad_{-k}$ are
particles of opposite momentum forming a pair that
constitutes the elementary excitations. Using the quadruple notation
\be
\label{quadruple3}
a^\mu=(a,\atd,\bd,\bt)\ ,\quad \xi^\mu=(\xi,\xtd,\ed,\et)\ ,
\ee
the preservation of CCR, $\dagger$ and $\sim$ gives
the conditions
\be
\ba{rcl}
B_4(\tau_3\otimes\tau_3)B_4^\dagger &=& (\tau_3\otimes\tau_3)\ ,\\
B_4(\id\otimes\tau_1) &=& (\id\otimes\tau_1)B_4\ .
\ea
\ee
Similarly to the case II a tensor product of a $SU(1,1)$
and a $SO(1,1)$
rotations would satisfy the conditions but we look for more
general solutions. For a real $B_4$ we find similarly
\be
B_4=\left(\ba{cc}
B_2(\theta_1)\cosh\phi &
B_2(\theta_2-\delta)\sinh\phi \\
B_2(\theta_1+\delta)\sinh\phi &
B_2(\theta_2)\cosh\phi \ea\right)\ ,
\ee
with the same interpretation as in case II. Again, we give the
free Hamiltonian
\bea
\hat{H}(\xi,\eta)&=&(\omega_a\cosh^2\phi+\omega_b\sinh^2\phi)
      (\xd\xi-\xtd\xt)+(\omega_b\cosh^2\phi+\omega_a\sinh^2\phi)
      (\ed\eta-\etd\et)\non\\
      & &+\,
      \frac{\sinh 2\phi}{2}(\omega_a+ \omega_b)\left\{
      \cosh(\theta_1-\theta_2+\delta)
      [\xd\ed+\eta\xi-\xtd\etd-\et\xt] \right. \non\\
      & &- \left.
      \sinh(\theta_1-\theta_2+\delta)
      [\xd\et+\etd\xi-\xtd\eta-\ed\xt]\right\}\ ,
\eea
and the number expectation values
\bea
\xvacl\ad a\xvacr &=& \cosh^2\phi\sinh^2\theta_1+
      \sinh^2\phi\cosh^2(\theta_2-\delta)\ ,\non\\
\xvacl\bd b\xvacr &=& \cosh^2\phi\sinh^2\theta_2+
      \sinh^2\phi\cosh^2(\theta_1+\delta)\ .
\eea
%
\section{Extension to fermions}
\seqnoll
\label{fbyffermi}
The analysis of Sec.\ref{genalpha} and Sec.\ref{fbyfbose}
can be repeated
for fermions. Since we want the fermions to anti--commute
we need to define the thermal doublet like $a^\mu=(a,i\atd)\ ,\
\ad^\mu=(\ad,-i\at)$ and we use the convention
$\tilde{\at}=a$\cite{Ojima81}.
Differences in signs in the canonical anti--commutation relation
(CAR) and the factors of $i$
lead to slightly different conditions compared
to Sec.\ref{fbyfbose}, but they can all
be solved in a similar manner.
%
\subsection{$2\times2$ Bogoliubov matrices}
We start with the extension of the thermal $2\times2$ Bogoliubov
transformation to the cases when Hermitian and
tilde conjugation are not preserved.
Defining $a^\mu=F_2^{\mu\nu}\xi^\nu$ and $\ad^\mu=G_2^{\mu\nu}\xd^\nu$
the CAR gives
\be
F_2^{-1}=G_2^T\ .
\ee
The only restriction on $F_2$ is that it should be
invertible. When we require $\sim$ invariance we
get the extra condition
\be
F_2\tau_2F^\dagger=\tau_2\ ,
\ee
just as for bosons. The condition for
preserving Hermiticity is however
\be
F_2F_2^\dagger=\id\ .
\ee
We summarize the various possibilities for fermions and bosons
in Tab.1.
\\ \\
\btab{|c||c|c|}   \hline
preserve            & $\sim$ \ yes & $\sim$ no
\\ \hline\hline
\btab{c} \ \\ $\dagger$ yes  \\ bosons \\ $\beta\in \real$ \\ \ \etab
&
$e^{i\theta}\left(\btab{cc} $\pm\sqrt{1+\beta^2}$ & $\beta$ \\
                 $\beta$ & $\pm\sqrt{1+\beta^2}$ \etab \right)$
&
$e^{i\theta}\left(\btab{cc} $\sqrt{1+\beta^2}e^{i\phi}$
            & $\beta e^{i\varphi}$ \\
              $\beta e^{-i\varphi}$ & $\sqrt{1+\beta^2}e^{-i\phi}$
            \etab \right)$
\\ \hline
\btab{c} \ \\ $\dagger$ yes  \\ fermions \\ $\alpha\in [-1,1]$ \\ \ \etab
&
$e^{i\theta}\left(\btab{cc} $\pm\sqrt{1-\alpha^2}$ & $\alpha$ \\
                 $-\alpha$ & $\pm\sqrt{1-\alpha^2}$ \etab \right)$
&
$e^{i\theta}\left(\btab{cc} $\sqrt{1-\alpha^2}e^{i\phi}$
            & $\alpha e^{i\varphi}$ \\
              $-\alpha e^{-i\varphi}$ & $\sqrt{1-\alpha^2}e^{-i\phi}$
            \etab \right)$
\\ \hline
\btab{c} \ \\ \ \\ $\dagger$ no \\ \ \\ \  \etab
&
$e^{i\theta}\left(\btab{cc} $\zeta$ & $\beta$ \\
                 $\gamma$ & $\frac{1+\beta\gamma}{\zeta}$ \etab
            \right) \in U(1)\times SL(2,\real)$
&
$\left(\btab{cc} $\zeta$ & $\beta$ \\
                 $\gamma$ & $\delta$ \etab
            \right) \in GL(2,\complex)$
\\ & & $\zeta\delta-\beta\gamma\neq 0$
\\ \hline
\etab
\\ \ \\  \ \bc Table 1. The $2\times2$ Bogoliubov matrices
for bosons and fermions.\ec \ \\
%
\subsection{$4\times4$ Bogoliubov matrices}
It is straightforward to determine what kind of
$4\times4$ Bogoliubov transformations are possible for
fermions along the lines of Sec.\ref{fbyfbose}.
\\ \\
{\bf Case I}
\\
The fermionic cases are quite different from the bosonic ones.
At zero temperature a transformation like $\xi=\alpha a+\beta\ad$
can only satisfy the CAR if $\alpha\beta=0$ (and
$\Abs{\alpha}^2+\Abs{\beta}^2=1$), i.e. if one of $\alpha$
and $\beta$ is zero.
This kind of constraint survives also at finite
temperature.

Using the quadruple notation
\be
a^\mu=(a,i\atd,\ad,-i\at)\ ,\quad \xi^\mu=(\xi,i\xtd,\xd,-i\xt)\ ,
\ee
the condition for preserving CAR, $\dagger$ and $\sim$ are
\be
\ba{rrcl}
{\rm CAR:} & F_4(\tau_1\otimes\id)F_4^T &=& (\tau_1\otimes\id)\ ,\\
\dagger: & F_4^*(\tau_1\otimes\id) &=& (\tau_1\otimes\id)F_4\ , \\
\sim   : & F_4^*(\tau_2\otimes\tau_2) &=& (\tau_2\otimes\tau_2)F_4\ .
\ea
\ee
The only solution which is continuously connected to the
identity is in fact
\be
\left(\ba{c} a \\ i\atd \ea\right) =
\left(\ba{rr} e^{i\psi_1}\cos\theta & e^{i\psi_2}\sin\theta \\
      -e^{i\psi_2}\sin\theta & e^{i\psi_1}\cos\theta \ea\right)
\left(\ba{c} \xi \\ i\xtd \ea\right)\ ,
\ee
and its Hermitian conjugate. We conclude that it is only
the usual thermal mixing that is possible for a single
fermionic mode.
\\ \\
In connection with the observation that there is no non--trivial
mixing of fermions as described above we want to
comment on another convention
for tilde conjugation of fermions. In many places in
the literature it is postulated that $\tilde{\at}=-a$
for fermions
and the thermal doublet $(a,\atd)$ is used. This notation gives the
same possible $2\times2$ Bogoliubov matrices at equilibrium,
and no physical difference between the two conventions is known
so far.
For the $4\times 4$ mixing considered here in case I the conditions
for preservation of CAR, $\dagger$-- and $\sim$--conjugation
allows for non--trivial transformations using this alternative
convention. To be more precise, if we define
\be
a=\alpha\xi+\beta\xtd+\gamma\xd+\delta\xt\ ,
\ee
and extend it through $\dagger$-- and the altered
$\sim$--conjugation,
the CAR can be satisfied even when all $\alpha,\ \beta,\
\gamma$ and $\delta$ are non--zero. For example we can take
\be
\ba{ll}
\alpha=\cos\phi\cos\theta \ , & \gamma=-\sin\phi\sin\theta\ ,\\
\beta=\cos\phi\sin\theta \ , & \delta=\sin\phi\cos\theta\ ,
\ea
\ee
which is not possible with the convention $\tilde{\at}=a$.
The relations between the different conventions
for non--equilibrium states is not yet clear
and we continue to use $\tilde{\at}=a$ since
the construction of such anti--commuting tilde
operators is possible from the basic Tomita-Takesaki
modular theory \cite{Ojima81}.
\\ \\
\pagebreak  \\
{\bf Case II}
\\
Mixing of two species of fermions is important e.g. for massive
neutrinos in connection with the solar neutrino problem.
When we write the quadruple like
\be
a^\mu=(a,i\atd,b,i\btd)\ ,\quad \xi^\mu=(\xi,i\xtd,\eta,i\etd)\ ,
\ee
the condition for CAR, $\dagger$ and $\sim$ are
\be
\label{fermi2cond}
\ba{rcl}
F_4F_4^\dagger &=&\id\otimes\id\ , \\
F_4(\id\otimes\tau_2) &=& (\id\otimes\tau_2)F_4\ .
\ea
\ee
A tensor product of the particle and thermal mixing
fulfills \Eqref{fermi2cond}, but more general
solutions are allowed. Assuming
that $F_4$ is real we get
\be
\label{F42}
F_4=\left(\ba{rr} F_2(\theta_1)\cos\phi &
F_2(\theta_2-\delta)\sin\phi
\\  -F_2(\theta_1+\delta)\sin\phi &
F_2(\theta_2)\cos\phi \ea\right)\ ,
\ee
where
\be
F_2(\theta)=\left(\ba{rr} \cos\theta & \sin\theta \\
                        -\sin\theta & \cos\theta
            \ea\right)\ .
\ee
The free Hamiltonian and the number expectation values
are in this case
\bea
\hat{H}(\xi,\eta)&=&
      (\omega_a\cos^2\phi+\omega_b\sin^2\phi)
      (\xd\xi-\xtd\xt)+
      (\omega_b\cos^2\phi+\omega_a\sin^2\phi)
      (\ed\eta-\etd\et)\non\\
      & &+\frac{\sin 2\phi}{2}(\omega_a-\omega_b)
      \left\{\cos(\theta_1-\theta_2+\delta)
      [\xd\eta+\ed\xi-\xtd\et-\etd\xt]\right. \non\\
      & &- \left.
      i\sin(\theta_1-\theta_2+\delta)
      [\xd\etd+\xtd\ed-\et\xi-\eta\xt]\right\}\ ,\\
\xvacl\ad a\xvacr &=& \cos^2\phi\sin^2\theta_1+
      \sin^2\phi\sin^2(\theta_2-\delta)\ ,\non\\
\xvacl\bd b\xvacr &=& \cos^2\phi\sin^2\theta_2+
      \sin^2\phi\sin^2(\theta_1+\delta)\ .
\eea
\\ \\
\pagebreak \\
{\bf Case III}
\\
In the BCS theory of superconductivity there is a mixing
between $a_{k,\uparrow}$ and $\ad_{-k,\downarrow}$.
Thermo field dynamics was first invented as an operator
formalism to deal with superconductivity at finite
temperature \cite{LeplaeUM74} and this application is
still of current interest \cite{Greenberg93}.
A general mixing of
the BCS type preserving CAR, $\dagger$--
and $\sim$--conjugation
must satisfy
\be
\ba{rcl}
F_4F_4^\dagger &=&\id\otimes\id\ , \\
F_4(\tau_3\otimes\tau_2)&=&(\tau_3\otimes\tau_2)F_4^*\ ,
\ea
\ee
if we use the notation $a^\mu=(a,i\atd,\bd,-i\bt)$.
The solution is
\be
F_4=\left(\ba{cc} F_2(\theta_1)\cos\phi &
\tau_3F_2(\theta_2-\delta)\sin\phi \\
-\tau_3F_2(\theta_1-\delta)\sin\phi & F_2(\theta_2)\cos\phi
\ea \right)\ .
\ee
The transformed Hamiltonian
and the number expectation values are
\bea
      \hat{H}(\xi,\eta)&=&
      (\omega_a\cos^2\phi-\omega_b\sin^2\phi)
      (\xd\xi-\xtd\xt)   +
      (\omega_b\cos^2\phi-\omega_a\sin^2\phi)
      (\ed\eta-\etd\et)\non\\
      & &+\frac{\sin 2\phi}{2}(\omega_a-\omega_b)\left\{
      \cos(\theta_1+\theta_2-\delta)
      [\xd\ed+\eta\xi-\xtd\etd-\et\xt] \right. \non\\
      & &-\left. i\sin(\theta_1+\theta_2-\delta)
      [\xd\et+\xtd\eta-\ed\xt-\etd\xi]\right\}\ ,\\
\xvacl\ad a\xvacr &=& \cos^2\phi\sin^2\theta_1+
      \sin^2\phi\cos^2(\theta_2-\delta)\ ,\non\\
\xvacl\bd b\xvacr &=& \cos^2\phi\sin^2\theta_2+
      \sin^2\phi\cos^2(\theta_1-\delta)\ .
\eea
%
\section{Conclusions}
\seqnoll
We have generalized the $\alpha$--degree of freedom
and discussed the relation to Hermitian and tilde
conjugation. Depending on which algebraic properties one
wishes to preserve when representing the
operators in the thermal Hilbert space
different parametrizations of the thermal Bogoliubov
transformation are possible. Some of the parameters have
been used earlier to simplify time dependent
non--equilibrium calculations in TFD and we hope
that the extensions considered here will also turn out
to be useful.

Mixing of modes occurs in several systems at zero temperature.
We have combined particle mixing and thermal mixing for
a number of cases for bosons and fermions.
The essential result is that the number of physical
parameters increases when we consider both mixings at
the same time. Even when Hermitian and tilde conjugation
are preserved, so that the corresponding $\alpha$--degree
of freedom is eliminated, there remains a new parameter
which we call $\delta$. One can, for instance, imagine
a system that
undergoes a phase transition acquires a value of $\delta$
which is determined by the dynamics. Quite generally we
would like to stress that the set of all Bogoliubov
transformations is very rich when one leaves the
simple equilibrium case, mostly considered in the
literature so far.

The meaning of the parameter $\delta$ has yet to
be clarified in greater detail but we can speculate
about its role in the thermal states. Let us take the
bosonic case II as an example and let $\theta_1=\theta_2$
for simplicity. The usual way of determining the value
of the mixing parameter $\phi$ is to diagonalize
the quadratic part of the interacting Hamiltonian
\cite{LeplaeUM74,Greenberg93}. In \Eqref{HamboseII}
we start out from a diagonal Hamiltonian but we can see
what kind of terms the non--zero $\delta$ generates.
The term with a factor $\cosh(\delta)$ contains no
mixing of tilde and non--tilde modes, they are
only in the $\sinh(\delta)$ term. The mixing of
tilde and non--tilde operators is usually related
to dissipation so we may expect that the $\delta$
parameter is also related to dissipation. These are
only speculations that have to be substantiated by
further research.
\\ \\
{\bf Acknowledgement}
\\
P. Elmfors wish to thank The Swedish Institute for
financial support and The Theoretical Physics Institute,
Edmonton, Alberta, for their hospitality during his stay.
We also thank Dr. T. Einarsson for providing a
Mathematica package for normal ordering.
%

%
%
%
%
\end{document}